\documentclass[11pt]{article}

\usepackage{amsfonts,amsmath,amssymb,amsthm}

\usepackage{hhline}
\usepackage{hyperref}
\usepackage[english]{babel}
\usepackage[utf8]{inputenc}
\usepackage[dvipsnames]{xcolor}
\usepackage{graphicx}
\usepackage[T1]{fontenc}
\usepackage{textcomp}
\usepackage{floatflt}
\usepackage[font={footnotesize}]{caption}
\usepackage{subcaption}
\usepackage{color}
\usepackage{makecell}
\usepackage{algpseudocode}
\usepackage{algorithm}
\usepackage{geometry}
\definecolor{szin}{rgb}{0,0.44,0.4}
\definecolor{szin2}{rgb}{0.902,0.2705,0}
\definecolor{szin3}{rgb}{0,0.5,0}
\hypersetup{pdfpagemode=UseNone, pdftoolbar=true, breaklinks=true,
colorlinks=true,
linkcolor=szin2,
linktoc=page,
citecolor=szin3,bookmarks=false}
\makeindex

\def\Pr{\mathbb{P}}

\newcommand{\TTT}{{\widehat{\mathcal{T}}}}

\def\L{\mathcal{L}}

\DeclareMathOperator\erf{erf}

\newcommand{\TT}{\mathcal{T}}

\def\1{\mathbf{1}}
\def\aa{\alpha}

\usepackage{mathrsfs}

\def\P{\Pr}

\def\md{\mid}
\def\Bb#1#2{{\def\md{\bigm| }#1\bigl(#2\bigr)}}

\def\Bs#1#2{{\def\md{\mid}#1(#2)}}
\def\Pb{\Bb\P}

\def\Ps{\Bs\P}

\newcommand{\avg}[1]{\left< #1 \right>} 

\graphicspath{{./plots/} }

\begin{document}


\title{Modelling structure and predicting dynamics of discussion threads in online boards}

\author{ Alexey N. Medvedev
\thanks{Namur Institute for Complex Networks (naXys), Universit\'e de Namur, Rempart de la Vierge, 8, Namur 5000 Belgium and ICTEAM, Universit\'e Catholique de Louvain, Av. Georges Lemaitre 4, 1348 Ottignies-Louvain-la-Neuve, Belgium. Email: {\tt an\_medvedev `at' yahoo.com}}
\and
Jean-Charles Delvenne
\thanks{ICTEAM and CORE, Universit\'{e} Catholique de Louvain, 1348 Louvain-la-Neuve, Belgium. Email: {\tt jean-charles.delvenne `at' uclouvain.be}}
\and
Renaud Lambiotte
\thanks{naXys, Universit\'{e} de Namur, 5000 Namur, Belgium, Mathematical Institute, University of Oxford, OX2 6GG Oxford, UK. Email: {\tt renaud.lambiotte `at' maths.ox.ac.uk}}
}

\date{}

\maketitle

\begin{abstract}
Internet boards are platforms for online discussion about a variety of topics. On these boards, individuals may start a new thread on a specific matter, or leave comments in an existing discussion. The resulting collective process leads to the formation of `discussion trees’, where nodes represent a post and comments, and an edge represents a ‘reply-to’ relation. The structure of discussion trees has been analysed in previous works, but mainly from a static perspective. 

In this paper, we focus on the structural and dynamical properties of discussion trees by modelling their formation as a self-exciting Hawkes process with a branching structure. We first study a Reddit dataset of all posts and comments submitted between 2008 and 2014 to show that the structure of the trees resemble those produced by a Galton-Watson process with a special root offspring distribution. The dynamical aspect of the model is then used to predict future commenting activity and the final size of a discussion tree. We compare the efficiency of our approach with previous works and show its superiority for the prediction of discussion dynamics. In particular we find that the Reddit discussion trees, being fundamentally broader than e.g. the Twitter cascades and developing with a qualitatively different dynamics, require appropriate models. The tree structure and the dynamics are influenced mainly by the age of the discussion, rather than by the number of acquired comments. The post and the comments are further commented by different processes, which is a key ingredient in modelling of discussions. 
\end{abstract}

\section{Introduction}

Online social media offer a rich source of information for the study of social behaviour. Depending on the platform, different types of tree-like cascading patterns emerge as a consequence of social interactions \cite{Adar2005,Gruhl2004}. For example, on Twitter or on Facebook, people interact via sharing or retweeting content, which may lead to large cascades of events \cite{Kwak2010,Dow2013,zhao2015seismic}. In email networks, people may forward messages to their acquaintances, resulting in cascading trees of email forwards \cite{IribarrenMoro2011}. In online boards like Digg or Reddit, people interact by discussing certain posts, which leads to the formation of reply cascades or, so-called,  `discussion trees' \cite{Kumar2010}. In general, researchers have been interested in two main questions: what is the shape of these cascades and what is the dynamics of their evolution?

Several works have focused on the structural properties of cascades and on the mechanisms that could generate them. In \cite{Gomez2011}, for instance, the authors considered cascades in four large Internet boards (Slashdot, Barrapunto, Meneame and Wikipedia) and proposed a generating model based on preferential attachment mechanism. In \cite{GomezLitvak2013}, the authors enriched this model by incorporating a notion of novelty to the comments, and aimed at reproducing the relation between the width and the depth of discussion trees. Note that these models focus exclusively on the structure of the trees  and not on the timings at which events take place. In \cite{WangHuberman2012}, in contrast, the authors introduce a  theoretical model for the structural and temporal evolution of discussions, based on the notion of L\'evy processes, but the mean-field nature of the model limits its calibration with real-world datasets, and thus its applicability. Up to our knowledge, the problem of predicting both structure and dynamics of discussions remains open.
 
The interplay between the structure and the dynamics of cascades has been considered in other social media data sets, for instance as a feature in a machine learning framework to predict successful cascades from unsuccessful ones \cite{cheng2014}. From a modelling perspective, word-of-mouth cascades in email networks have been modelled by Bellman-Harris branching processes with lognormal inter-branching times \cite{IribarrenMoro2011}. Hawkes processes were successfully applied to study and predict the evolution of retweet cascades in Twitter \cite{Perc2017, zhao2015seismic, kobayashi2016tideh}. The structure of so-called `reply trees' was also studied by Nishi et. al in \cite{nishi2016reply}, where authors proposed simple principles for the formation of such trees. It is important to stress that the mechanisms behind retweet cascades and reply trees on one side, and discussion trees on the other side, are intrinsically different, due to the different designs of their platforms. The former is strongly affected by the structure of an underlying social network, for instance the followee-follower network of Twitter \cite{PercPONE2017}, while the latter makes content equally available to any visitor of a website, principally based on recency. Posts with little attention get quickly drowned in the stream of newer or exciting material.

In this paper, we analyse the dataset of all comments to all posts in Reddit initiated in the period between Jan 2008 and Jan 2015 and concentrate on `discussion trees' formed for each post. We consider discussion trees as a random object, thus aggregating all trees over all subreddits. Our main contribution is the introduction of a model for the growth of discussion trees based on Hawkes processes and its validation on the dataset. The model incorporates both structure and dynamics, and is used for predicting the future commenting activity. The model suggests that discussion trees in Reddit structurally resemble branching trees with a special root offspring distribution, and the commenting dynamics is well predicted by self-exciting process compared to contemporary models. The paper is organized as follows. In Section~\ref{sec:dataset}, we present and describe the main properties of the dataset, in Section~\ref{sec:model} we present the model with parameters estimation procedure, in Section~\ref{sec:evaluation} we set up the evaluation procedure and present the existing models that we use as baselines for comparison, in Section~\ref{sec:results} we present the results and we conclude the paper with the discussion in Section~\ref{sec:discussion}.

\section{Dataset}\label{sec:dataset}
Reddit is a platform designed to share and discuss  content online.  Since  its creation in 2005, when it presented itself as the "frontpage of the Internet", the platform has grown and diversified  into a global social system with its own list of rules, coalitions and even language \cite{Strohmaier2014}. The main driving force of Reddit is its users, who act by submitting posts made of graphic or textual material and by participating in discussions attached to these posts. Reddit is further divided into subforums, or subreddits, which can be set up by their users \cite{Tan2015}. 

Reddit is organised as follows: users post a message on a particular subject, which appears on the website and becomes available to all users. Each post has a section for comments, where users write a reply to the post or to previous comments of this post. In this paper, we analyze the aggregated dataset of all posts and comments submitted to Reddit between Jan, 2008 and Jan, 2015 \cite{dataset_reddit, dataset_link}. Comments form a discussion, which can be represented as a \textit{rooted tree}, where the root is a designated node representing the post itself and each other node represents a comment. There is a link between two nodes if there is a `reply-to' relation between them. We disregard the edge direction as it may be recovered from the timestamps or tree level distance from the root. Each post and comment contains a unique id and information about its author, its content, the creation timestamp and a link to the node to which it replies. In the following, we disregard the information on the content of posts and comments, and concentrate mainly on the temporal and structural properties of the trees. 

\begin{figure}[t]
\centering
\includegraphics[width=0.8\linewidth]{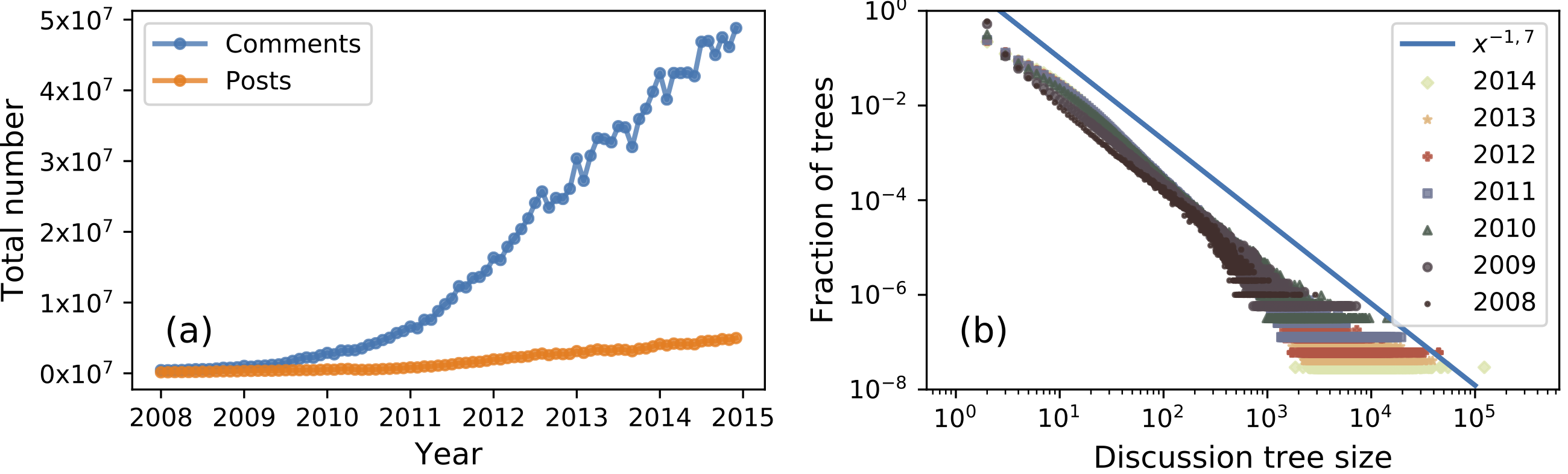}
\caption{(a) Total monthly submissions to Reddit: the orange curve represents posts and the blue curve represents comments. (b) Size distribution of discussion trees, collected for each year from 2008 to 2014. The power-law  $\sim x^{-1.7}$, in blue, is a guide for an eye.}
\label{fig:post_comment_stats}
\end{figure}

The dataset in total contains more than 150 million posts and around 1.4 billion comments. The stable growth of the number of total monthly submissions is presented on Figure~\ref{fig:post_comment_stats}, (a). The growth rate of both posts and comments increases between 2010 and 2012, and remains roughly stable afterwards, which is probably related to the release of a mobile app for smartphones \cite{mobile_app}. While the total number of discussion trees grows in time, their size distribution remains approximately the same (see Figure~\ref{fig:post_comment_stats}, (b)), with a shape close to a power-law $\Ps{\xi>x}\sim x^{-\aa}$ with  $\aa=1.7$.

\begin{figure}[t]
\centering
\includegraphics[width=0.75\linewidth]{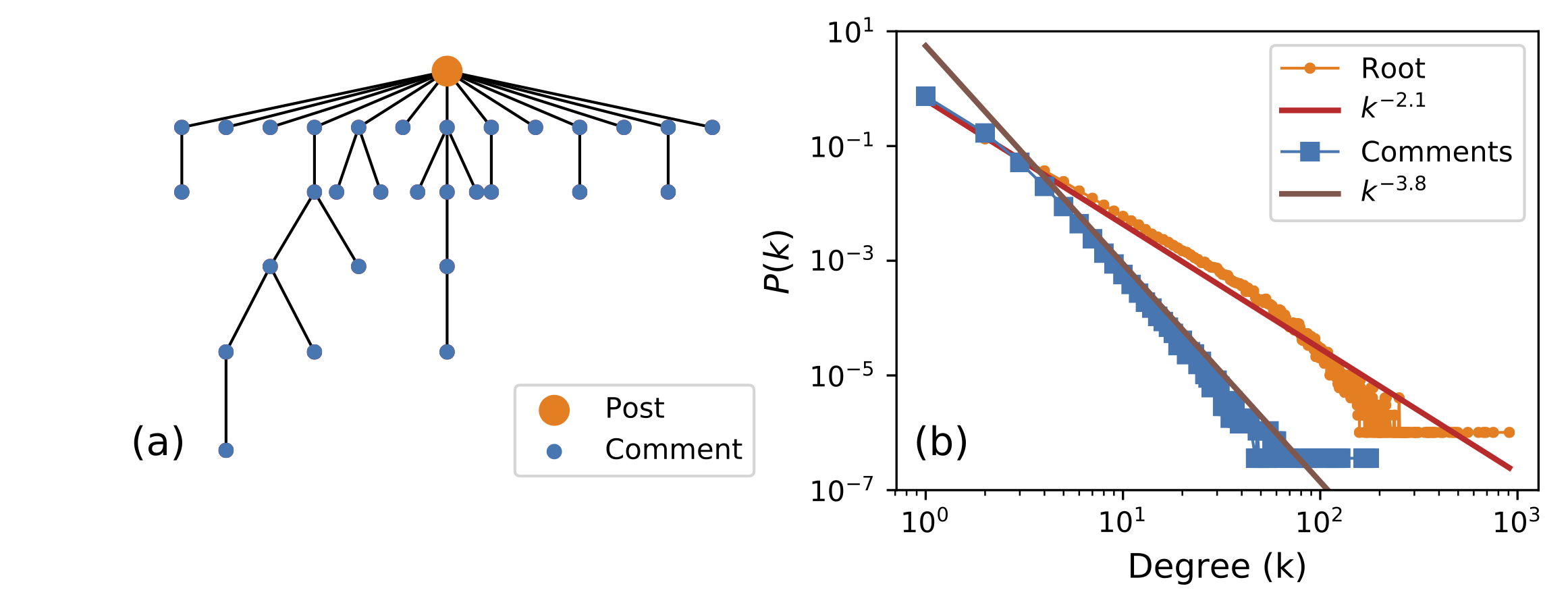}
\caption{(a) Example of a discussion tree. The root is coloured in red and depicted with a larger sized node. (b) Degree distribution for the roots and for comments in 2008. Roots have a significantly larger degree, on average, than comments. The distributions are well approximated by a power-law with different exponents.}
\label{fig:degree_distribution}
\end{figure}

The Reddit platform is designed such that users are first exposed to the content and general information of a post, while comments are only visible after clicking under the post. This difference implies that the root has a larger degree, on average, than a comment, as shown in Figure~\ref{fig:degree_distribution}, (a), an observation already made in other Internet boards \cite{Gomez2011} and for reply trees in Twitter \cite{nishi2016reply}. This observation has important consequences, as it suggests that the degree of the root is determined by a different process than for other nodes. This conclusion is confirmed in Figure~\ref{fig:degree_distribution}, (b), where we show the degree distribution of the root (in red) and the forward degree distribution of all other vertices (in blue) in 2008 are clearly different. By definition, the \textit{forward degree} is  the degree of a comment minus one, and it corresponds to  the number of comments made in reply to a certain comment. Although a slight tail cutoff is visible for the root degree distribution, both distributions are well approximated by a power-law with exponents close to $2.1$ for the root and $3.8$ for the comments. Note that the forward degree distribution for comments does not appear to change with the distance to from root. Similar distributions were obtained for other years.

\begin{figure}[t]
\centering
\includegraphics[width=0.8\linewidth]{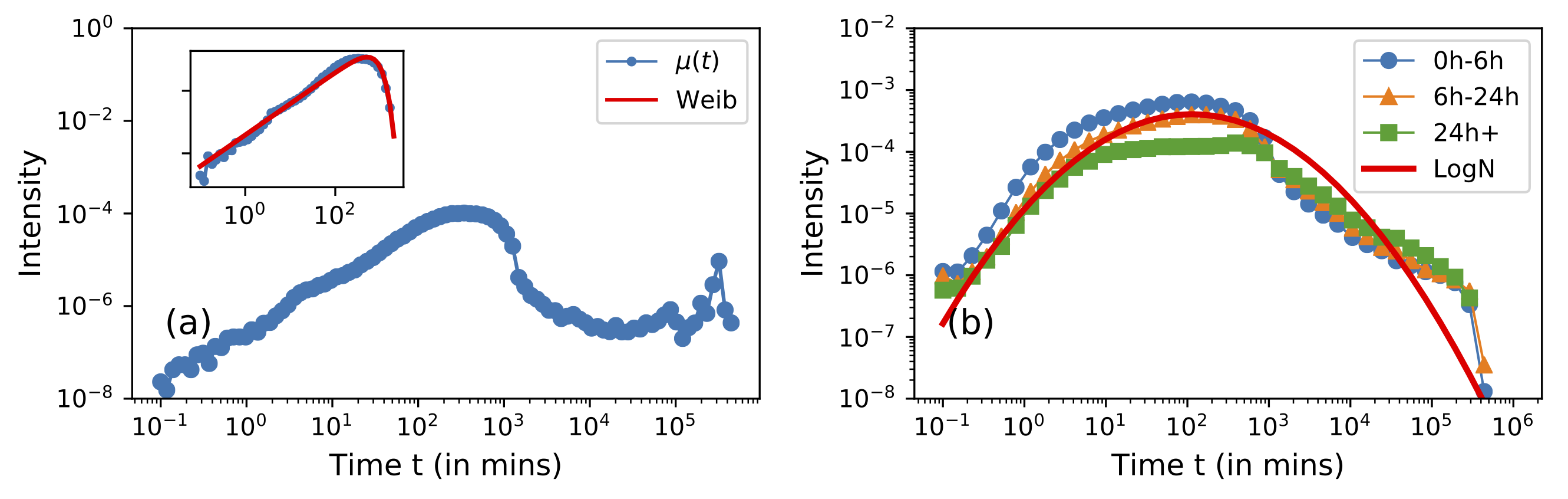}
\caption{Response time distributions for (a) the root and (b) comments. The inset of (a) depicts the intensity of response times for the roots within 36 hours, along with the Weibull pdf fit given with the red curve. The response times for comments on (b) are divided into three sets: \textit{early} comments that appeared within 6 hours from post's appearance, \textit{mid} comments, created within 6 and 24 hours after the post's creation and the remaining \textit{late} comments. The red line corresponds to a lognormal distribution with parameters $\mu=7.9$ and $\sigma=1.78$.}
\label{fig:response_times}
\end{figure}

Another argument supporting the evidence of the different nature of comments and posts can be built from the temporal characteristics of the trees. Consider the discussion tree $\TT$ with node set $V=\{v_1,\dots,v_n\}$, where $v_1=s$ is the root. Denote as $\tau_{v_i}\geqslant 0$ the timestamp of the node $v_i$, where $1\leqslant i\leqslant n$. We further assume that $\tau_{v_1}=0$, as it is possible to make an offset by the posting time. Consider a node $v$ and its forward neighbour $u$. We call the difference $\tau_{v}-\tau_{u}$ the \textit{response time} of the node $v$ to the node $u$, namely the node to which $v$ replies to. The data shows that the response time distribution follows a different pattern for the root and for the comments (see Figure~\ref{fig:response_times}). In particular, the distribution for the root within 36 hours  shows an initial power-law increase  followed by an exponential decay (see inset to Figure~\ref{fig:response_times}, (a)), which can be well fitted by a pdf of a Weibull distribution:
$$\mathsf{Weib}(a,b,\aa) = a(\alpha/b)\left(t/b\right)^{\aa-1}\exp\left(-\left(t/b\right)^{\aa}\right),$$
where $a>0,\,b>0,\,\aa>0$. Parameter $a$ can be interpreted as a mean number of offsprings of the root, $b$ is a timescale parameter and $\aa$ governs the shape of distribution. Weibull distributions have been used in survival analysis and extreme value theory, as they represent a limit distribution of a minimum of $n$ i.i.d. random variables, satisfying a specific criteria \cite{vdHofstad2016}. We observe from the Figure~\ref{fig:response_times}, (a), that the tail of the response time distribution does not follow any specific pattern, which may be the consequence of aggregation of many posts with different parameters $a$ and $b$.

To test whether the age of a comment affects how triggering new comments, we divide the comments into three subsets: 1) \textit{early comments} that appeared within 6 hours, 2) \textit{mid comments}, created within 6 and 24 hours  and 3) the remaining \textit{late comments}. According to the division we have about $0.5\%$ of total comments classified as late and about $37\%$ as mid. In each case, as we notice on Figure~\ref{fig:response_times}, (b), due to the absence of abrupt exponential decay for large times, the response time distribution is well fitted by a lognormal distribution:
$$\mathsf{LogN}(\mu,\sigma) = \frac{1}{t\sigma\sqrt{2\pi}}\exp\left(-\frac{(\log t - \mu)^2}{2\sigma^2}\right),$$
where $\mu,\sigma>0$. The lognormal distribution naturally emerges  when considering the product of i.i.d. random variables with finite variance (see e.g. \cite{feller1968}) and it has been used in a variety of problems, for instance as a kernel to model  citation dynamics \cite{Wang2013, Shen2014}. We also investigated whether the level of the comment may have an impact on its response time, but no significant influence in the data was found.

\begin{figure}[t]
\centering
\includegraphics[width=1\linewidth]{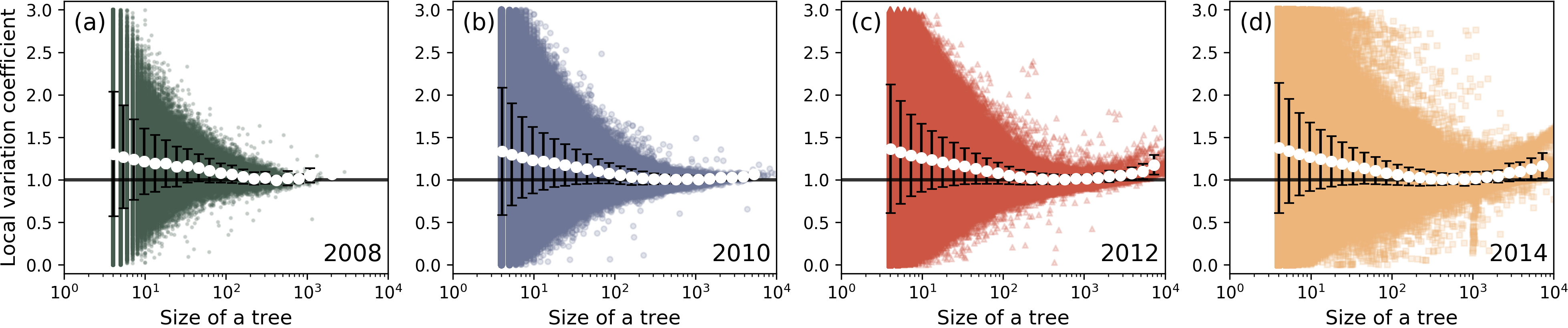}
\caption{$LV$ scores for discussion trees in the dataset for : (a) 2008, (b) 2010, (c) 2012, (d) 2014. The scatter plot of individual scores is accompanied with the mean score  in white and bars for the standard deviation. It is observed that  the size of discussion trees increases in time, however the overall pattern of the distribution of $LV$ coefficients remains similar. For each year, we present the tree sizes in the range from 1 to 10000 nodes, and as the tree size grows, the $LV$ coefficients tend to be closer 1 on average (given as white circles) and show less dispersion as well (bars represent standard deviation).}
\label{fig:local_variation}
\end{figure}

To get a further insight into the temporal characteristics of the dataset, we  consider, for each discussion, the time series $\tau$ of creation times of each comment, defined as an ordered ascending collection of timings $\tau=\{\tau_1<\tau_2<\dots <\tau_n\}$. Due to its non-stationarity, we characterise correlations in the time series by its  \textit{local variation coefficient} $LV$ \cite{Shinomoto2003, Shinomoto2005, Miura2006, Shinomoto2011, LambiotteCeyda2015}. The local variation coefficient is defined by comparing temporal variations with their local rates 
\begin{equation}\label{eq.lv}
LV(\tau)=\frac{3}{n-1}\sum\limits_{i=1}^{n-1}\left(\frac{\delta \tau_{i+1}-\delta \tau_{i}}{\delta \tau_{i+1}+\delta \tau_{i}}\right)^2,
\end{equation}
where $\delta \tau_i = \tau_i - \tau_{i-1}$, and is thus specifically designed for non-stationary processes. 
The coefficient takes values in the interval $(0,3)$. $LV(\tau)$ is equal to 1 when the point process that generates $\tau$ is an inhomogeneous Poisson process. Deviations from 1 originate from local correlations in the underlying time series, either under the form of pairwise correlations between successive inter-event time intervals, e.g. $\delta\tau_{i+1}$ and $\delta\tau_i$, which tend to decrease $LV$, or because the inter-event time distribution is non-exponential \cite{LambiotteCeyda2015}. In Figure~\ref{fig:local_variation}, we observe a  relation between the size of a discussion tree and its value of $LV$, as larger trees tend to show values closer  to those of Poisson process. We will provide an explanation for this observation in Section~\ref{sec:results}. 

\section{Hawkes model of discussion trees}\label{sec:model}

A discussion tree may grow in two ways, either by adding comments to the original post or by adding a comment to an existing comment. As we showed, the temporal and structural properties of each mechanism appears to be different. We incorporate these findings in a self-exciting Hawkes process, a well-known point process where the intensity (i.e. rate function) evolves as
\begin{equation}\label{eq:hawkes}
\lambda(t) = \mu(t) + n_b\sum\limits_{i:\tau_i<t}\phi(t-\tau_i),
\end{equation}
and where $\mu(t)$ is the background intensity, $n_b$ is the branching number and $\phi(t)$ is the memory kernel. The process is self-exciting because the rate increases after the realisation of an event $i$ at time $\tau_i$ with a kernel $\phi(t-\tau_i)$ \cite{Hawkes1971}. By definition, the memory kernel satisfies the condition that $\avg{\phi}=\int\limits_0^{\infty}\phi(t)dt=1$ and it can be interpreted as the probability that a new triggered event takes place at time $t-\tau_i$. The branching number $n_b$ controls the amount of self-excitation and is equal to the average number of events directly triggered by event $i$. We also assume $\avg{\mu(t)}<\infty$, thus the background process dies out eventually. 

\begin{figure}[t]
\centering
\includegraphics[width=0.7\linewidth]{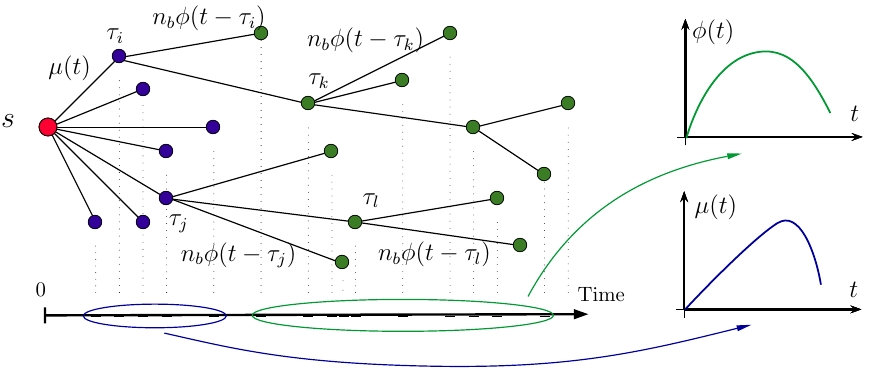}
\caption{Graphical description of a Hawkes process as a branching process.}
\label{fig:hawkes_description}
\end{figure}

The dynamics of this process can be equivalently described as a branching process. Let the root be created at time $t=0$ and start generating offsprings with a rate $\mu(t)$. Each offspring $i$, generated at time $\tau_i$, starts generating its own offsprings, with a rate $n_b\phi(t-\tau_i)$ (see Figure~\ref{fig:hawkes_description}). The resulting tree is a Galton-Watson tree with special root offspring distribution. The tree is almost surely finite if $n_b\leqslant 1$ \cite{ross1996, daley2003}. 

Hawkes processes have been successfully applied to model many real-world phenomena, including social interactions and information propagation in Twitter, gang rivalries over time \cite{Egesdal2010}, earthquakes and their aftershocks \cite{Ogata1998} and neuronal spikes of activity \cite{NeuronHawkes2010,NeuronHawkes2013}. Commonly used memory kernels have an exponential and power-law functional forms, describing short and long-range memory \cite{zipkin2016}. The process definition may also be adjusted to have a multiplicative effect of self-excitation, e.g.
\begin{equation}\label{eq:hawkes_2}
\bar{\lambda}(t) = \mu(t) \cdot \left(\sum\limits_{i:\tau_i<t}\phi_i(t-\tau_i)\right),
\end{equation} 
where $\mu(t)$ is a background intensity and $\phi_i(t)$ is a memory kernel of $i$'th event.  Hawkes processes with intensity in \eqref{eq:hawkes_2} have, in particular, been applied to predict Twitter cascades \cite{zhao2015seismic, kobayashi2016tideh}. However, this definition does not allow different mechanims for the root versus later comments, which motivates the use of  \eqref{eq:hawkes} for discussion trees.

Based on our empirical observations, we specify the functional forms of the kernels as follows: $\mu(t)$ is given by a Weibull pdf $\mathsf{Weib}(a,b,\aa)$, and the memory kernel $\phi(t)$ takes the form of a lognormal distribution $\mathsf{LogN}(\mu,\sigma)$.  $n_b$ is for the average number of offsprings of comments. 

The parameters of the model are estimated for each discussion tree separately using a maximum likelihood estimation (MLE). Since data contains information on the structure of discussions, we can also separately estimate parameters for the root process and the comments. The general formula for the maximum likelihood function of a non-homogeneous Poisson process with intensity $\lambda(t)$ is given by
$$\log\L(\tau_1,\tau_2,\dots,\tau_k|\theta)=-\int\limits_{0}^{\tau_k}\lambda(t)dt + \int\limits_{0}^{\tau_k}\log (\lambda(t))dN(t),$$
where $\tau_1,\dots, \tau_k$ are event times, $\theta$ is the set of parameters and $N(t)$ is the counting process \cite{ross1996,daley2003}. In our setting the loglikelihood function for the intensity $\mu(t)$ is 
\begin{equation}
\begin{split}
\log\L(\theta(\mu(t))|\tau) =& -a(1-\exp\left(-\left(\tau_k/b\right)^{\aa}\right)-\sum\limits_{i=1}^{k}\left(\left(\tau_k/b\right)^{\aa}+(\aa-1)\log(\tau_i)\right)+\\
&+k(\log(a)+\log(\aa)-\aa\log(b)).
\end{split}
\end{equation}
The loglikelihood function for $\phi(t)$ is given as follows
\begin{equation}
\begin{split}
\log\L(\theta(\phi(t))|\tau) =&-\frac{1}{2}\left(1+\erf\left(\frac{\log \tau_k-\mu}{\sqrt{2}\sigma}\right)\right) +k\log\left(\frac{1}{\sigma\sqrt{2\pi}}\right) \\
&- \sum\limits_{i=1}^{k}\left(\frac{(\log \tau_i-\mu)^2}{2\sigma^2}+\log \tau_i\right).
\end{split}
\end{equation}
Maximization of $\log\L(\theta|\tau) $ is performed using the L-BFGS-B algorithm, where parameters are constrained to be positive \cite{Zhu1997}. 

In order to estimate the branching number $n_b$ of a given tree, we collect the forward degrees of all comments across the tree and compute the average value:
$$n_b=\frac{1}{n-1}\sum_{v\neq root}(d_v-1) = 1-\frac{d_{root}}{n-1},$$
where $d_v$ is the degree of a node $v$ and $d_{root}$ is the degree of the root, 

\section{Model comparison}\label{sec:evaluation}
In this section, we provide an overview of the baseline models and evaluation metrics for model comparison. We evaluate the performance of our model on four sets of discussion trees in Reddit divided by years: (a) 2008, (b) 2010, (c) 2012 and (d) 2014. From each set, we take a subset of all discussion trees of \textit{small} size between 50 and 200 nodes and take a random selection of around 8000 trees in each subset. A similar procedure is performed for subsets of \textit{larger} discussion trees of size between 200 and 2000. The following division is used to show similarity of statistical properties of small and large trees.

\subsection{Modelling structure}

To our knowledge, the only models proposed for the structure of discussion trees are based on the preferrential attachment principle. In our analysis we use the \textit{preferential attachment (PA) growth model}, introduced in \cite{Gomez2011}. Note that this model does not provide information about the temporal evolution of the trees. In this model, trees grow with preferential attachment at discrete steps, as newly arriving nodes connect with a higher probability to the root or to a large degree node. The thread is thus viewed as a growing tree $\TT_n$ with $n$ nodes $\{1,2,\dots,n\}$, where a newly arriving node is attached to an existing node $k$ with probability
$$\Pb{\text{attach to node }k\md \TT_n} = \frac{1}{Z_n}(\beta_k d_{k,n})^{\gamma_k},$$
and where $d_{k,n}$ is the degree of node $k$ in $\TT_n$, $(\beta_k,\gamma_k) = (\beta, \gamma_c)$, if $k$ is the root and $(\beta_k,\gamma_k) = (1, \gamma)$ otherwise, and $Z_n$ is a normalizing factor for probability distribution. Parameters $\gamma_c, \gamma$ govern the strength of preferential attachment and $\beta$ stands for the root bias, which are fitted for a given tree by  maximizing the likelihood of arrival of each new node. 

We compare the quality of our model with respect to PA by performing Monte Carlo simulations. For each sample tree, we estimate the parameters of the Hawkes model and the PA model and then simulate 50 different trees from each of these models. The PA model takes the number of nodes as an input, thus we let each simulated PA tree have the same number of nodes as the Hawkes tree from the same simulation run for a proper comparison. A similar strategy was suggested in \cite{WangHuberman2012}. The difference in structure between the given tree $\TT$ and a generated one $\TTT$ is compared using the \textit{error per distance layer}
$$\varepsilon_d^{\min}(\TTT,\TT) = \frac{1}{d_{\min}}\sum_{k=1}^{d_{\min}}|\widehat{N_k} - N_k|,\quad \varepsilon_d^{\max}(\TTT,\TT)=\frac{1}{d_{\max}}\sum_{k=1}^{d_{\max}}|\widehat{N_k} - N_k|,$$
where $N_k$ (resp. $\widehat{N_k}$) is the number of nodes at distance $k$ from the root in $\TT$ (resp.  $\TTT$) and $d_{\min}$ (resp. $d_{\max}$) is the minimal (resp. maximal) depth of the trees $\TT$ and $\TTT$. These errors measure the mean deviation of the node distribution across the distance layers from the root between a sample tree $\TT$ and a simulated tree $\TTT$. The simulated tree may have depth different from a sample tree, therefore we record both errors. Simulation averages of the minimal $\varepsilon_d^{\min}$ and maximal $\varepsilon_d^{\max}$ errors over the number of runs are further recorded for each sample tree. 

\begin{figure}[t]
\centering
\begin{minipage}[b]{0.4\linewidth}
\centering
\includegraphics[width=0.99\linewidth]{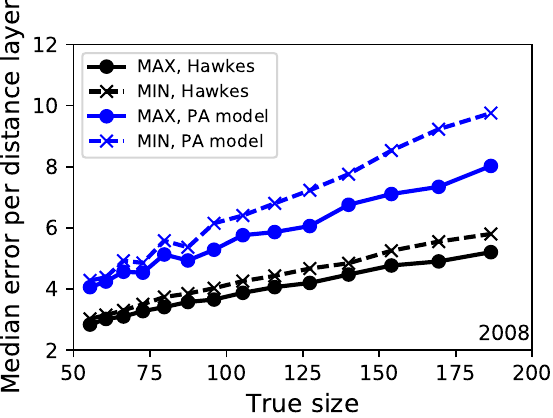}
\end{minipage}
\begin{minipage}[b]{0.4\linewidth}
\centering
\includegraphics[width=0.99\linewidth]{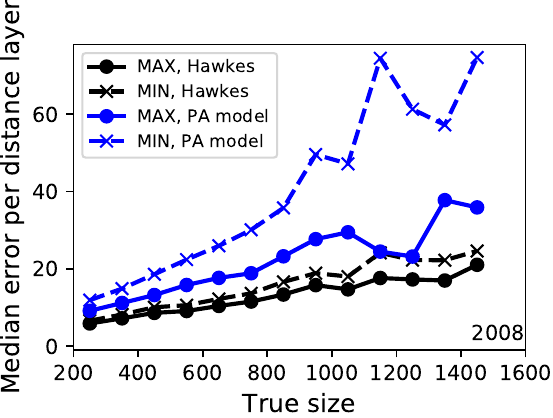}
\end{minipage}
\caption{Comparison of average errors in capturing the structure of discussion trees with Hawkes and PA models. The sample trees are binned into logarithmically sized bins and median errors for each bin are plotted. The errors are plotted for a sample of small (a) and large (b) trees in year 2008. MAX error corresponds to the simulation average $\avg{\varepsilon_d^{\max}}$ and MIN stands for the simulation average $\avg{\varepsilon_d^{\min}}$.}
\label{fig:results_distance_layer}
\end{figure}

\subsection{Predicting activity} 

Due to the lack of an existing model for the temporal behaviour of discussions, we consider two models previously designed for Twitter cascades and citation activity with special choice of activity kernels. The evaluation is performed as follows: after fitting the model parameters by observing the timing of events in a learning time window, the task is to predict its future evolution. Parameters evaluation depends on the observation time of the tree evolution. The tree is observed from the appearance of the root (post) at time $t=0$ up to the time $t_{learn}$ and the parameters are estimated from that truncated data. In our study, we use four different learning periods $t_{learn}$: 4, 6, 8 or 12 hours. The data shows that, after 12 hours, there remains on average less than $20\%$ of activity to predict, thus longer observation windows include almost all the discussion (see insets to Figure~\ref{fig:results_loglikelihood_2014} (d) and (h)). After learning the parameters, the fitness of the time series of the future temporal activity is evaluated through its loglikelihood score in a particular model (standard likelihood ratio test). We also use Monte-Carlo simulations to determine the average size error of the discussion prediction. As above, after 50 runs of simulation for each model, we calculate the average predicted total size $\avg{\hat{s}}$ of the discussion tree and compute the \textit{absolute relative size error} 
$$\varepsilon_s^{t_{learn}} = \frac{|\avg{\hat{s}}-s|}{s},$$
where $s$ is the size of a sample tree and $\hat{s}$ is the predicted size. The following models were used for comparison.

\textit{Dynamic Poisson model (DP) \cite{Crane2008,Agarwal2009}.} The time series is modelled as a point process with deterministic time-dependent intensity $\lambda(t)$. DP has previously been applied to Youtube views and Twitter cascades with a power-law functional form for the intensity. In our experiments on discussion trees, we found that the most accurate results were achieved with a lognormal intensity $\mathsf{LogN}(\mu,\sigma)$. The parameters are fitted using the maximum likelihood method. 

\textit{Reinforced Poisson Process (RPP) \cite{Shen2014,RP2015}.} The time series is generated by a point process with reinforced intensity $\lambda(t,k) = cf(t)r(k)$, where $t$ is time and $k=k(t)$ is the current number of events in the time series. The background intensity $f(t)$ models the general interest in the subject and was assumed to have a lognormal $\mathsf{LogN}(\mu,\sigma)$ functional form. The same kernel function has been used to model the citation dynamics \cite{Shen2014}. $r(k)=\sum_{i=1}^{k} \exp(-d i)$, where $d>0$, is the reinforcement factor, meaning that each new event $i$ adds an $\exp(-d i)$ increase to the overall intensity. 

We took these models in order to underline the complexity of the discussion process. If the DP models discussions uniformly without extra triggering, the RPP enriches the dynamics with time-independent reinforcement. The Hawkes process presents a next step of refinement, where reinforcement is itself time-dependent. We note the existence of more sophisticated models that use multiplicative Hawkes processes for prediction of Twitter cascades \cite{zhao2015seismic, kobayashi2016tideh}. However, the model \cite{zhao2015seismic} is only designed to predict the total size of the cascade, whereas we concentrate on total future activity, and \cite{kobayashi2016tideh} is designed to model circadian patterns, which are not present in our data. 

\section{Results}\label{sec:results}

\begin{figure}[t]
\centering
\includegraphics[width=0.97\linewidth]{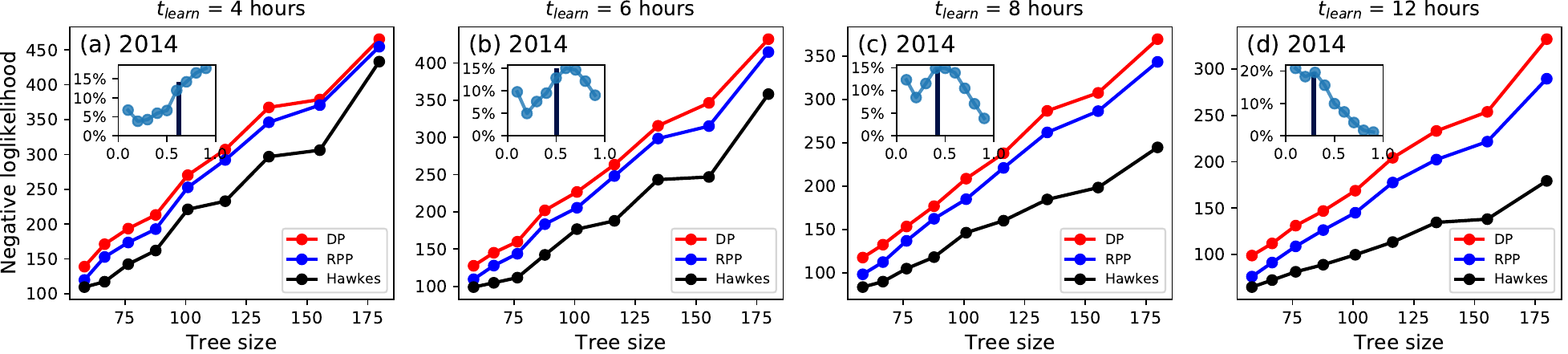}
\end{figure}
\begin{figure}[t]
\centering
\includegraphics[width=0.97\linewidth]{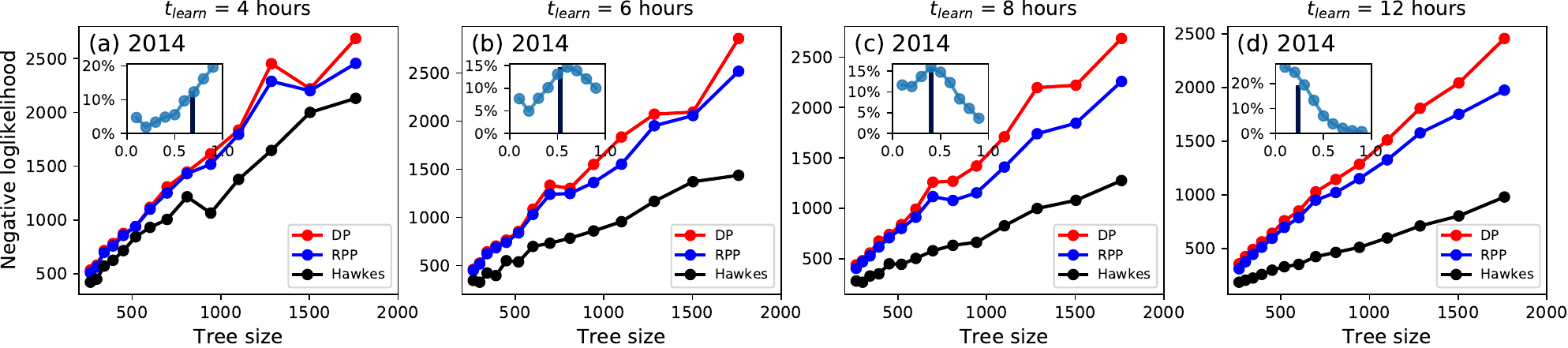}
\caption{Comparison of activity prediction performance of the Hawkes model with the models previously used for similar social interactions for one sample of small and one sample of large trees in 2014. The prediction performance is measured using the likelihood of the future activity in a sample tree beyond the learning time window. The learning time window $t_{learn}$ is available for the parameters evaluation, and is set to 4 ((a), (e)), 6 ((b), (f)), 8 ((c), (g)) and 12 ((d), (h)) hours. The sample trees are binned into logarithmically sized bins and median negative loglikelihoods are plotted for each bin. The insets show the distribution of the fraction of activity that remains to predict after the learning phase over the whole sample.}
\label{fig:results_loglikelihood_2014}
\end{figure}

The results for the structural modelling are shown in Figure~\ref{fig:results_distance_layer}. Due to the high computational complexity of the inference method of the PA model, we focus only on the year 2008. To present the results, we divide the discussion trees by size into logarithmically sized bins to obtain a more uniform number of samples in each bin. The median error is computed for each of the bins. The plot shows that despite both $\avg{\varepsilon_d^{\min}}$ and $\avg{\varepsilon_d^{\max}}$ grow with the discussion size, the Hawkes model outperforms the PA model in reproducing the tree shape. Although for small trees the results are comparable (see Figure~\ref{fig:results_distance_layer}, (a)), for large trees the gap between $\avg{\varepsilon_d^{\max}}$ errors tend to grow drastically (see Figure~\ref{fig:results_distance_layer}, (b)).

The results for the temporal prediction are illustrated for 2014 (see Appendix for other years). Again, we divide the trees into logarithmically sized bins to obtain a more uniform number of samples in each bin and calculate the median negative loglikelihood score for each bin.  We consistently observe a better fit between the remaining time series and the Hawkes model in the case of small trees, for all considered observation windows $t_{learn}$ (see Figure~\ref{fig:results_loglikelihood_2014}, (a)-(d)). In the insets, we observe that for $t_{learn}=4$ hours, only $35\%$ of total activity is used for the training. The Hawkes model already shows better fit of the data, however due to the lack of sufficient information, all three models show rather similar performance scores. Longer observation windows result in better parameters estimation and the Hawkes model takes the unbeatable lead for learning times larger than $t_{learn}=8$ hours when, on average, around $30\%$ of the cascade is yet to be predicted. 

For large trees, a sufficient performance increase is achieved already for $t_{learn} = 6$ hours (see Figure~\ref{fig:results_loglikelihood_2014}, (e)-(h)). We observe in the insets that in both cases the methods have to predict a similar fraction of activity. Therefore we can conclude that the active commenting period is similar for any post regardless of its size. Larger discussions are thus associated with a larger density of events within the discussions' life. Since the LV coefficient \eqref{eq.lv} for large discussions comes closer to 1, on average, one could hypothesize that bursts and self-excitation become less apparent, and the temporal process approaches a Poisson process. In general, assigning a specific process to the root is an essential ingredient for the prediction. 
\begin{figure}[t]
\centering
\begin{minipage}[b]{0.245\linewidth}
\centering
\includegraphics[width=1\linewidth]{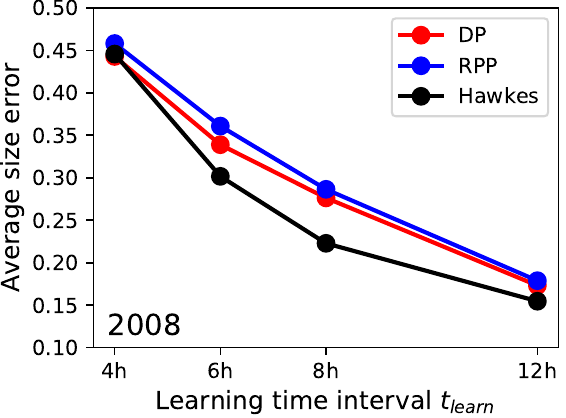}
\end{minipage}
\begin{minipage}[b]{0.245\linewidth}
\centering
\includegraphics[width=1\linewidth]{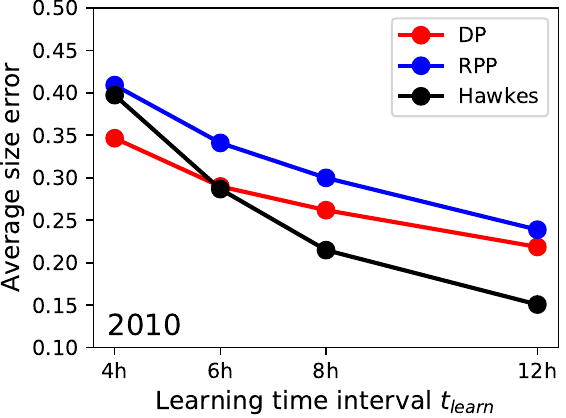}
\end{minipage}
\begin{minipage}[b]{0.245\linewidth}
\centering
\includegraphics[width=1\linewidth]{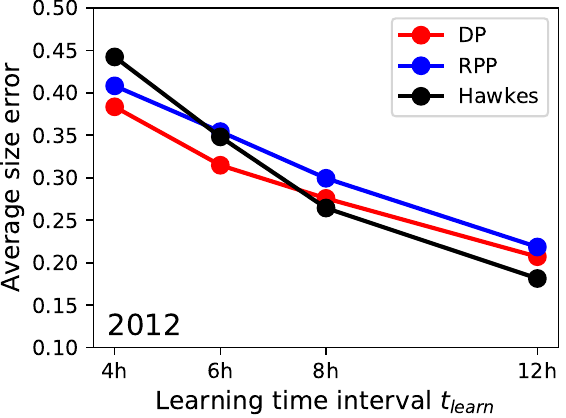}
\end{minipage}
\begin{minipage}[b]{0.245\linewidth}
\centering
\includegraphics[width=1\linewidth]{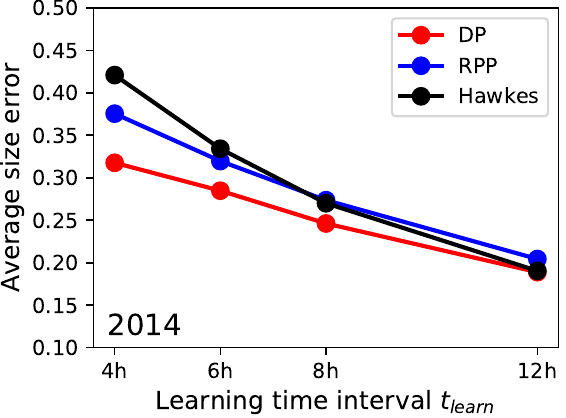}
\end{minipage}
\caption{Absolute relative size errors in predicting the size of discussions by Hawkes, DP and RPP models. The errors are represented for samples of small trees in years (a) 2008, (b) 2010, (c) 2012 and (d) 2014. Median relative size errors are displayed for the observed values of $\varepsilon_s^{t_{learn}}$.}
\label{fig:results_size_error}
\end{figure}

We show the relative error of total discussion size for samples of small trees in Figure \ref{fig:results_size_error}. For each observation window $t_{learn}$ we sample average relative errors and plot a median of these values. The short observation window produces larger size prediction error on average, however when $t_{learn}$ increases, the median error for all models steadily decreases. Although the Hawkes model shows comparable results in total size prediction, it is less accurate for small observation windows, when the lack of information complicates the estimation of a relatively large number of parameters. This limitation could, in principle, be overcome by implementing more sophisticated estimation procedures.

\section{Discussion}\label{sec:discussion}
Online boards play an important role for the exchange of ideas and the occurrence of debates in the online world. In this work, we have proposed a model allowing to reproduce and to predict the structural and temporal properties of discussion trees as they are observed in Reddit. It is important to note here that discussion trees are known to be system-specific and to depend on the user interface and system design of the online board. For example, reply trees observed in Twitter \cite{nishi2016reply} have different structural properties than those observed in this work. It was shown that a significant fraction of reply trees of size $n$ looks like a path of size $O(n)$ with minor branchings, whereas in case of branching trees the depth is bounded by $O(\sqrt{n})$. Retweet dynamics is essentially different from replies, suggesting a lognormal kernel instead of a power-law. This difference may arise from the different nature of the systems, as a comment in a discussion requires time, and produces new content, whereas retweeting is fast and requires essentially one click. Similarly, changing how information is presented to users on the website, e.g. most recent versus most popular, with or without a recommender system based on the user's interests, etc., is expected to result in different tree structures \cite{Aragon2017}. Nonetheless, the generality of our model and the fact that Hawkes-type processes have been successfully applied in a variety of social media suggests that our approach is robust.

As we have discussed, the main advantage of our model is the possibility to predict together, and with a good accuracy, the temporal and structural properties of discussion trees. For each dimension independently, however, some models might provide a better accuracy. For instance, for the modelling of  temporal properties, our results have shown that small discussions are better fitted by the Hawkes model than by an ordinary Poisson process. It is evident from Figure~\ref{fig:local_variation} that the LV scores for small trees exhibit a larger deviation from Poisson process. However, we also observed that for both small and large trees, the average duration of a post activity is comparable. In large trees, the overall presence of more branches leads to more condensed and more random time series, which may explain the decrease in accuracy of the Hawkes model. 

This work opens different venues for future research. For instance, one could improve the prediction power of the model by incorporating the content of posts and comments. It may specifically help to determine the current mood of the discussion and make a conclusion on its future direction \cite{Wikiwars2012}. Another way to measure the  popularity of a post is by using its rating. As each post and comment can be rated with `likes' and `dislikes', one could model the dynamics of conflicting debates but also the score of a post, defined as its number of `likes' minus its number of `dislikes', as a predictor for its future popularity \cite{Leskovec2010}. Collection of a dataset with timestamped `likes' and `dislikes' would also provide a deeper insight on the dynamics of opinion formation.

\section*{Funding}
This work was supported by Concerted Research Action (ARC) supported by the Federation Wallonia-Brussels Contract ARC 14/19-060; Flagship European Research Area Network (FLAG-ERA) Joint Transnational Call “FuturICT 2.0” (J.-C.D. and R.L.); and the Russian Foundation of Basic Research [grant number 16-01-00499] (A.M.)

\section*{Acknowledgement}
The authors thank Leto Peel and Daniele Cassese for fruitful discussions and helpful suggestions and anonymous reviewers for their useful comments that helped to improve the quality of the manuscript.

\bibliographystyle{imaiai}
\bibliography{literature} 
\newpage
\section{Appendix}
\subsection{Activity prediction: small trees}
\begin{figure}[ht]
\centering
\includegraphics[width=0.95\linewidth]{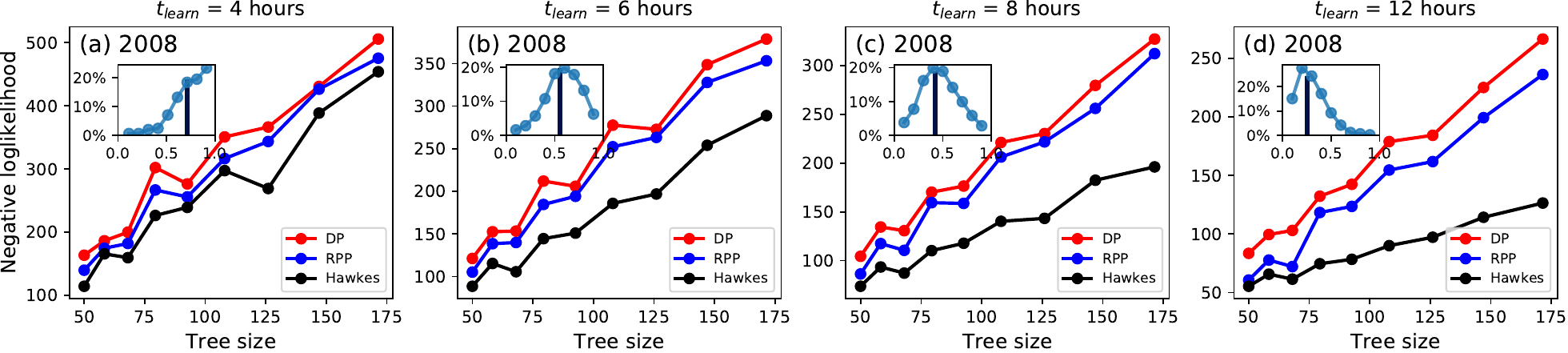}
\vspace{5mm}

\includegraphics[width=0.95\linewidth]{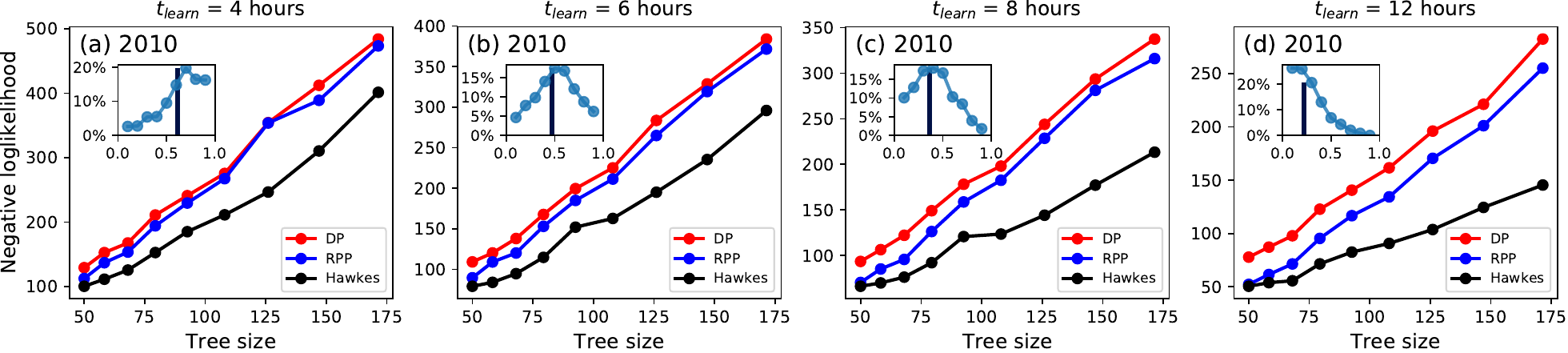}
\vspace{5mm}

\includegraphics[width=0.95\linewidth]{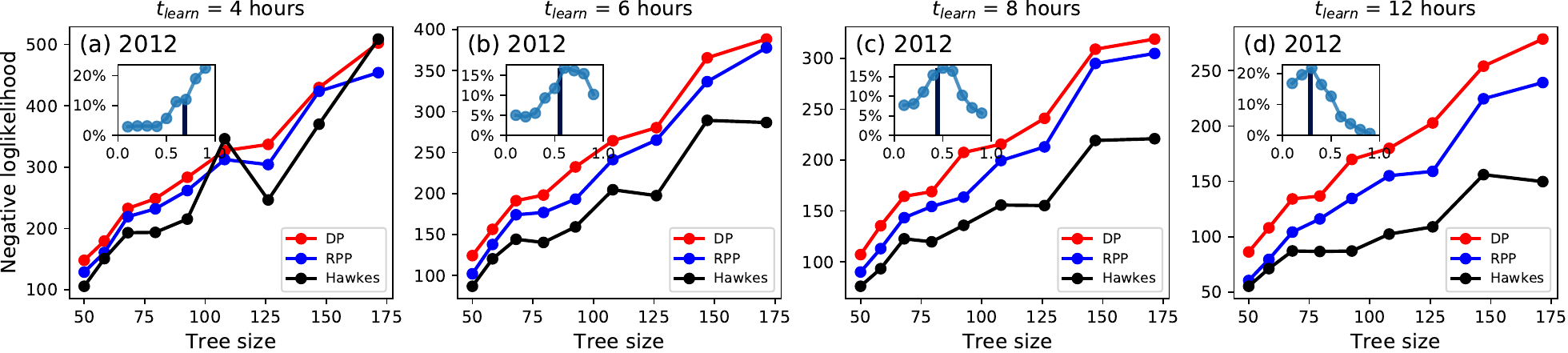}
\caption{
Performance of the activity prediction for Hawkes, DP and RPP models, for one sample of small trees in year 2008, 2010 and 2012. Predictive performance is measured using the likelihood of the future activity in a sample tree beyond the learning time window. The learning time window $t_{learn}$ is available for the parameters evaluation, and is set to 4 ((a), (e), (i)), 6 ((b), (f), (j)), 8 ((c), (g), (k)) and 12 ((d), (h), (l)) hours. The sample trees are binned into logarithmically sized bins and median negative loglikelihoods are plotted for each bin. The insets show the distribution of the fraction of activity that remains to predict after the learning phase over the whole sample.}
\label{fig:results_loglikelihood_2012_small}
\end{figure}

\newpage
\subsection{Activity prediction: large trees}
\begin{figure}[ht]
\centering
\includegraphics[width=0.95\linewidth]{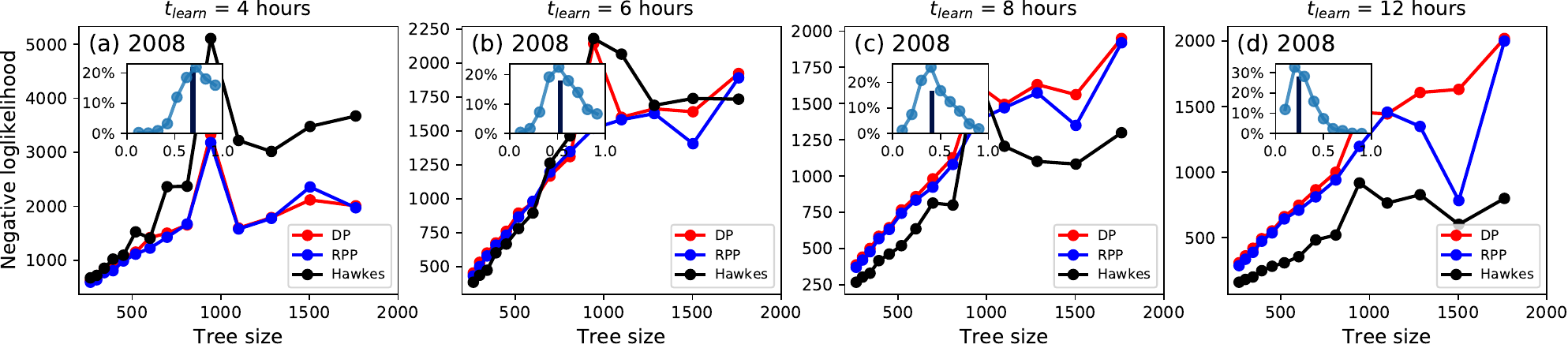}
\vspace{5mm}

\includegraphics[width=0.95\linewidth]{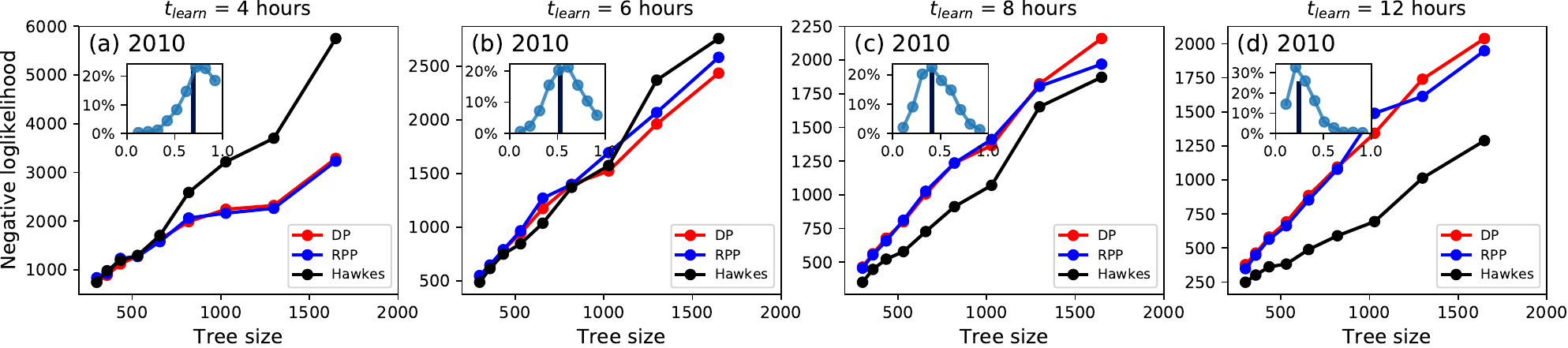}
\vspace{5mm}

\includegraphics[width=0.95\linewidth]{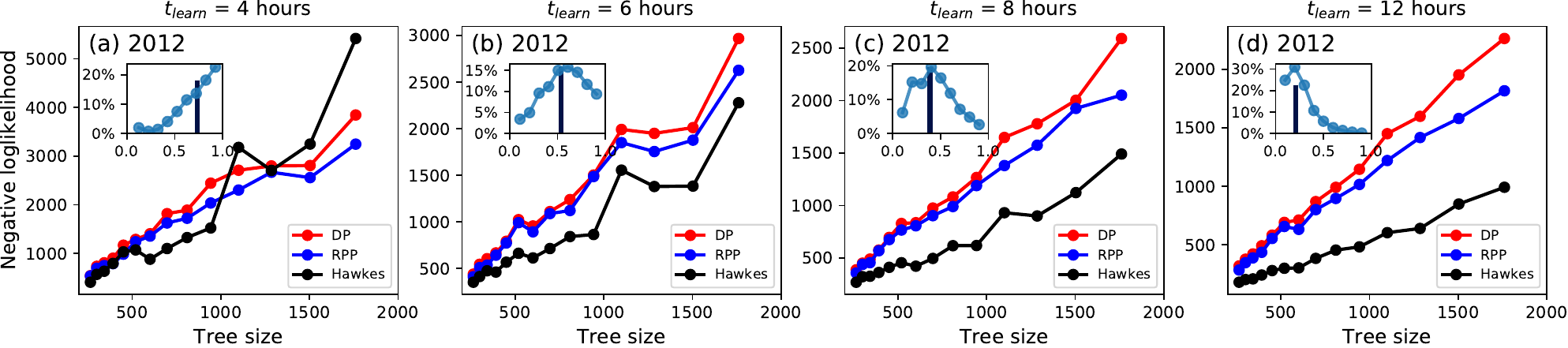}
\caption{
Performance of the activity prediction for Hawkes, DP and RPP models, for one sample of large trees in year 2008, 2010 and 2012. Predictive performance is measured using the likelihood of the future activity in a sample tree beyond the learning time window. The learning time window $t_{learn}$ is available for the parameters evaluation, and is set to 4 ((a), (e), (i)), 6 ((b), (f), (j)), 8 ((c), (g), (k)) and 12 ((d), (h), (l)) hours. The sample trees are binned into logarithmically sized bins and median negative loglikelihoods are plotted for each bin. The insets show the distribution of the fraction of activity that remains to predict after the learning phase over the whole sample.}
\label{fig:results_loglikelihood_2012_large}
\end{figure}
\end{document}